\documentclass[%
 reprint,
 amsmath,amssymb,
 aps,
nofootinbib,
superscriptaddress
]{revtex4-2}

\usepackage{graphicx}
\usepackage{dcolumn}
\usepackage{bm}
\usepackage[colorlinks=true]{hyperref}
\usepackage{ulem}

\usepackage{footnote}
\usepackage{xcolor}

\usepackage{caption}
\usepackage{subcaption}
\usepackage{amssymb}

\begin{document}

\title{Measuring deviations from the Kerr geometry with black hole ringdown}

\author{Kallol Dey}
\affiliation{School of Physics, Indian Institute of Science Education and Research Thiruvananthapuram, Maruthamala PO, Vithura, Thiruvananthapuram 695551, Kerala, India}

\author{Enrico Barausse}
\affiliation{SISSA, Via Bonomea 265, 34136 Trieste, Italy and INFN Sezione di Trieste}
\affiliation{IFPU - Institute for Fundamental Physics of the Universe, Via Beirut 2, 34014 Trieste, Italy}

\author{Soumen Basak}
\affiliation{School of Physics, Indian Institute of Science Education and Research Thiruvananthapuram, Maruthamala PO, Vithura, Thiruvananthapuram 695551, Kerala, India}

\begin{abstract}
Black holes in  General Relativity are famously characterized  by two ``hairs'' only, the mass and the spin of the Kerr spacetime. Theories extending General Relativity, however, allow in principle for additional black hole charges, which will generally modify the multipole structure of the Kerr solution. Here, we show that gravitational wave observations of the post-merger ringdown signal from black hole binaries may permit measuring these additional ``hairs''. We do so by considering spacetime geometries differing from the Kerr one at the level of the quadrupole moment, and computing the differences of their quasinormal mode frequencies  from the Kerr ones in the eikonal limit. We then perform a Bayesian analysis with current and future gravitational wave data and compute posterior constraints for the quadrupole deviation away from Kerr. We find that the inclusion of higher modes, which are potentially observable by future detectors, will allow for constraining deviations from the Kerr quadrupole at percent level.

\end{abstract}

\maketitle

\section{Introduction}\label{sec:introduction}
The most generic stationary and asymptotically flat black hole (BH) spacetimes of General Relativity (GR) are given by the Kerr-Newman solution~\cite{Kerr:1963ud, Newman:1965my, Carter:1968rr,PhysRevLett.34.905}, which is characterized by three parameters (mass, spin and electric charge). Since the electric charge is currently believed to be astrophysically unimportant~\cite{Barausse:2014tra}, astrophysical BHs are expected to possess only two charges (or ``hairs''), i.e. the mass and the spin. This is usually referred to as the ``no-hair theorem'' of GR.
Theories extending GR, however, may allow for BHs different from the Kerr (or Kerr-Newman) geometry~\cite{Barausse:2008xv}. These BH spacetimes will generally present additional ``hairs'', see e.g.~\cite{Berti:2015itd} for a review. As a result, they will present a  multipole structure different from that of Kerr BHs, from which they may deviate already at the quadrupole order (or higher).  

Gravitational wave (GW) observations of extreme mass-ratio inspirals have long been understood to hold the potential to test these putative deviations from Kerr~\cite{Ryan_1995,Ryan_1997}, with projected bounds on the quadrupole deviation with the Laser Interferometer Space Antenna (LISA)~\cite{LISA:2017pwj} reaching a fractional order of $10^{-5}$~\cite{Babak:2017tow}. Similar constraints on deviations  from Kerr can also be obtained with X-ray observations of accretion disks around BHs (either ones of the continuum spectrum~\cite{Bambi_2011_2,Bambi_2011_1} or  fluorescent iron lines~\cite{Ni:2016uik, Dabrowski:2000qv})
or, more recently, with the Event Horizon Telescope~\cite{eht_rezzolla,eht_sgra}. Moreover, with the growing number of GW signals detected by the LIGO-Virgo-KAGRA (LVK) collaboration~\cite{LIGOScientific:2016aoc, LIGOScientific:2017bnn, LIGOScientific:2017ycc, LIGOScientific:2017vwq, LIGOScientific:2017vox, LIGOScientific:2020aai, LIGOScientific:2020stg, LIGOScientific:2020zkf, LIGOScientific:2020iuh}, attempts have also been made at exploiting the  inspiral phase of  stellar-origin BH binaries for similar tests of the no-hair theorem~\cite{Barausse:2016eii,Cardenas-Avendano:2019zxd,Carson:2020iik, LIGOScientific:2019fpa, LIGOScientific:2020tif, LIGOScientific:2021sio}.

Tests of GR with the post-merger ringdown signal,  known as ``BH spectroscopy'', also have a long history, dating back to~\cite{1980ApJ...239..292D,Dreyer:2003bv,Berti:2005ys}. These tests hinge on the fact that within GR the (complex) quasinormal mode (QNM) frequencies are only a function of mass and spin (again due to the no-hair theorem), a hypothesis that can be tested by measuring two independent modes. The LVK detections are so far in agreement with the QNM ringdown frequencies predicted by GR, with evidence for the presence of the dominant $\ell=m=2 $, $n=0$ mode (in GW150914) and possibly\footnote{These claims 
have been shown to depend on the characterization of the detector noise, the data analysis methods, the choice of starting time for the ringdown phase, and even nonlinear
effects in the ringdown modeling~\cite{Cotesta:2022pci,Finch:2022ynt,Isi:2022mhy,Capano:2022zqm,Sberna:2021eui,Cheung:2022rbm,Mitman:2022qdl}.} also the $\ell=m=3 $, $n=0$ (in GW190521) and  $\ell=m=2 $, $n=1$   (in GW150914) modes. However, tests of GR with QNMs require higher signal-to-noise ratio, which will only be possible with next generation detectors~\cite{Berti:2016lat,Cabero:2019zyt,Ota:2021ypb,Bhagwat:2021kwv}. In the meantime, much work has already gone into developing theory agnostic parametrizations of the deviations of the ringdown signal from the GR prediction~\cite{Meidam:2014jpa,Maselli:2019mjd,Cardoso:2019mqo,McManus:2019ulj,Volkel:2020daa,Konoplya:2022pbc,Volkel:2022aca,Volkel:2022khh,Franchini:2022axs}.

In this paper, we will study what constraints on the deviation of the quadrupole moment away from the Kerr spacetime's value can be obtained from the ringdown signal of the GW150914  event. We will also show how these constraints will improve with next-generation detectors such the Einstein Telescope. In order to describe the non-Kerr BH spacetime, we will utilize two metric ansatze commonly employed in the literature on tests of the quadrupole moment of the Kerr metric, i.e. the
Manko-Novikov (MN) metric~\cite{Manko_1992} and the Johannsen-Psaltis (JP) metric~\cite{Psaltis_2011}.

The former is a vacuum stationary and axisymmetric solution of the Einstein equations of GR, which can describe the exterior of a rotating star or exotic compact object (see e.g. \cite{Berti:2003nb}), but which is also defined in the strong field region. In that regime, however, pathologies necessarily appear (because of the no-hair theorem of GR), i.e. 
the spacetime presents a partial event horizon with a curvature singularity on the equatorial plane, as well as closed timelike curves. These pathologies are ``covered'' by the star's or exotic compact object's matter if the metric describes one such body. The MN metric allows for deviations from the Kerr geometry at any multipole order, but in this work we will only study quadrupole deviations.
The JP metric, instead, is not a solution of the GR field equations, nor  of those of any extensions of GR. It is a purely phenomenological parametrization of  possible deviations from the Kerr hypothesis (see also \cite{Rezzolla:2014mua,Konoplya:2016jvv,Siqueira:2022tbc,Allahyari:2019umx,Allahyari:2018cmg} for a similar metric ansatze),
and it has been widely used for tests of the no-hair theorem with electromagnetic observations~\cite{Kong:2014wha, Bambi:2011ek, Bambi:2012pa, Bambi:2015ldr}.

Starting from either of these two geometries, we compute the deviations of the QNM frequencies of the dominant $\ell=m=2$, $n=0$ mode from the Kerr values, in the eikonal (i.e. geometric optics) approximation. Although this approximation is formally only valid in the limit of large $\ell$, it has been shown to approximately describe also the low-$\ell$ modes in GR, and it has also been applied beyond GR~\cite{Glampedakis_2017, Carson:2020iik, Glampedakis:2019dqh}.

We then use Bayesian methods to obtain posterior constraints on the quadrupole moment's deviation from Kerr, using existing ringdown data (GW150914) and future simulated ones (assuming detections by the Einstein Telescope (ET)). 
We find that the posteriors for the quadrupole deviation obtained from GW150914 are consistent with Kerr. Similar systems detected by ET will allow us to further improve this constraint, and break parameter degeneracies that appear between the remnant mass, spin and the quadrupole deviation.

This paper is organized as follows: In section \ref{sec:metric-intro}, we discuss the MN and the JP metric ansatze. In section \ref{sec:eikonal} we describe the eikonal approximation technique for the calculation of QNMs. The analysis techniques adopted in this paper are introduced in section \ref{sec:methodology}, followed by a discussion of the results in section \ref{sec:results}. We explore how future detectors might effect this analysis in section \ref{sec:future-detectors}.

\section{Quadrupole deviation from Kerr}\label{sec:metric-intro}

In the most general case, the gravitational field around a GR compact object can be described by a stationary, axisymmetric and asymptotically flat vacuum spacetime. The corresponding metric can be expressed in terms of the Geroch-Hansen moments~\cite{Hansen:1974zz, Geroch:1970cd} {$M_\ell, S_\ell$}, where $M_\ell$ and $S_\ell$ are respectively the mass and current moments. It can be shown that the non-zero mass moments are respectively the mass $M_0=M$, the mass quadrupole $M_2=Q$, and higher order even moments, and the non-zero current moments are the angular momentum $S_1=J$, $S_3$ and higher order odd moments. In the case of a Kerr spacetime, the following relation  connects $M_\ell$ and $S_\ell$
\begin{equation}\label{eq:nohair}
    M_\ell + i S_\ell = M\left(i\frac{J}{M}\right)^\ell
\end{equation}
This is the celebrated `no hair' theorem. However, if the  compact object is not a Kerr BH, Eq.~\eqref{eq:nohair} may not hold true and the quadrupole moment $Q$ may deviate from its expected Kerr value of $-a^2 M$. In this paper, we consider two geometries (the MN and JP ones) that describe compact objects with quadrupole moments different from Kerr.

\subsection{Manko Novikov spacetime}\label{sec:MNmetric}

The MN metric \cite{Manko_1992} describes a stationary, axisymmetric and asymptotically flat vacuum solution of Einstein's equations with arbitrary mass multipoles moments. In its full form, it has an infinite number of free parameters, but it can also reduce to the Kerr metric under appropriate conditions. The MN metric can be written in prolate spheroidal coordinates as~\cite{Bambi_2011_1, Bambi_2011_2}
\begin{widetext}
\begin{equation}\label{eq:MN1}
    ds^2 = -f(dt-\omega d\phi)^2 + \frac{k^2 e^{2\gamma}}{f}(x^2 - y^2)\left(\frac{dx^2}{x^2 - 1} + \frac{dy^2}{1 - y^2}\right)
    + \frac{k^2}{f}\left(x^2 - 1\right) \left(1 - y^2\right) d\phi^2
\end{equation}
where 
\begin{equation}\label{eq:MN2}
    f = e^{2\psi} \frac{A}{B} \,, \quad \omega = 2k e^{-2\psi}\frac{C}{A} - \frac{4k\alpha}{1-\alpha^2} \,, \quad e^{2\gamma}= e^{2\gamma'}\frac{A}{\left(x^2-1\right)\left(1-\alpha^2\right)^2}\,,\quad  \psi = \sum_{n=1}^{+\infty} \frac{\alpha_n P_n}{R^{n+1}}\,,
\end{equation}
\begin{subequations}\label{eq:MN3}
    \begin{eqnarray}
        \gamma' =&& \frac{1}{2}\text{log}\frac{x^2-1}{x^2-y^2} + \sum_{m,n=1}^{+\infty} \frac{(m+1)(n+1)\alpha_m\alpha_n}{(m+n+2)R^{m+n+2}} \left(P_{m+1}P_{n+1} - P_m P_n\right)\nonumber \\
        &&+  \sum_{n=1}^{+\infty}\alpha_n \left(\left(-1\right)^{n+1} - 1 + \sum_{k=0}^n \frac{x-y+\left(-1\right)^{n-k}\left(x+y\right)}{R^{k+1}}P_k\right) \,,
    \end{eqnarray}
    \begin{equation}
        A = \left(x^2-1\right)\left(1+\tilde{a}\tilde{b}\right)^2 - \left(1-y^2\right)\left(\tilde{b}-\tilde{a}\right)^2\,,
    \end{equation}
    \begin{equation}
        B = \left\{x+1+\left(x-1\right)\tilde{a}\tilde{b}\right\}^2 + \left\{\left(1+y\right)\tilde{a}+\left(1-y\right)\tilde{b}\right\}^2\,,
    \end{equation}
    \begin{equation}
        C = \left(x^2-1\right)\left(1+\tilde{a}\tilde{b}\right)\left\{\tilde{b}-\tilde{a}-y\left(\tilde{a}+\tilde{b}\right)\right\} + \left(1-y^2\right)\left(\tilde{b}-\tilde{a}\right)\left\{1+\tilde{a}\tilde{b}+x\left(1-\tilde{a}\tilde{b}\right)\right\}\,,
    \end{equation}
    \begin{equation}
        \tilde{a} = -\alpha \exp{\Bigg[\sum_{n=1}^{+\infty}2\alpha_n\left(1-\sum_{k=0}^n\frac{x-y}{R^{k+1} P_k}\right)\Bigg]}\,,
    \end{equation}
    \begin{equation}
        \tilde{b} = \alpha \exp{\Bigg[\sum_{n=1}^{+\infty} 2\alpha_n \left(\left(-1\right)^n+\sum_{k=0}^n\frac{\left(-1\right)^{n-k+1}\left(x+y\right)}{R^{k+1}}P_k\right)\Bigg]}\,.
    \end{equation}
\end{subequations}
\end{widetext}
where $R=\sqrt{x^2+y^2-1}$ and $P_n$ are Legendre polynomials such that
\begin{align}
    P_n &= P_n\left(\frac{xy}{R}\right)\,,\\
    P_n(\zeta) &= \frac{1}{2^n n!}\frac{d^n}{d\zeta^n}\left(\zeta^2 - 1\right)^n\,.
\end{align}
These can also be expressed in  Boyer-Lindquist coordinates $\left(r,\theta\right)$ using
\begin{equation}
    r=kx+M, \qquad \cos{\theta}=y
\end{equation}
We note that the metric used in this paper is the same as in~\cite{Bambi_2011_2}, which fixed some typos in the original MN metric~\cite{Manko_1992}.
Here, $k$ is related to  the mass of the spacetime, $\alpha$ to the spin and $\alpha_n$ to the mass multipole moments, indexed by $n = 1$ (dipole), $n=2$ (quadrupole), and so on. The MN metric reduces to the Schwarzschild metric when $\alpha=0$ and $\alpha_n=0$, and to the Kerr metric when $\alpha_n=0$.

In its most general form, the MN metric is not free from pathologies. This is expected because the only asymptotically flat and stationary BH solution of Einstein's vacuum equations that is non-singular on and outside the event horizon  is the Kerr metric (no hair theorem). In fact, the event horizon of the MN metric lies at $x=1$, and it has (in general) a naked singularity on the equatorial plane (at $x=1, y=0$) \cite{Manko_1992, Bambi_2011_1}. It is however possible that this curvature singularity will not
exist, as it may be covered by the exotic compact object's matter, whose gravitational field will only be described by the MN metric in the exterior.

In this study, we will not  consider the MN metric in its general form. Instead, we will focus on the special case of $\alpha_n=0$ for all $n\neq 2$. This corresponds to Kerr with an additional quadrupole deviation. Accordingly, the non-zero parameters in the MN metric are $k$, $\alpha$ and $\alpha_2$, which are related to the mass $M$ and dimensionless spin parameter $\chi$ by
\begin{align}
    k &= M \frac{1-\alpha^2}{1+\alpha^2}\,,\\
    \alpha &= \frac{\sqrt{1-\chi^2}-1}{\chi}\,,\\
    \alpha_2 &= q \frac{M^3}{k^3}\,,
\end{align}
where $q$ is the anomalous quadrupole moment, given by $q = -(Q-Q_K)/M^3$, with $Q$ and $Q_K$ respectively the MN and   Kerr quadrupole moments. 

This MN metric is interesting because it allows one to perform null tests of the no-hair theorem, by measuring $q$. In more detail,
$q=0$ corresponds to Kerr, $q>0$ would imply that the compact object under consideration is more oblate than a Kerr BH, while $q<0$ would point at a more prolate object. A statistically significant measurement of a non-zero value of $q$ would signal a violation of the no-hair theorem. Such a use of the MN metric for inference using GWs was proposed e.g. in \cite{Gair_2008}. This metric has also been used in studies trying to constrain quadrupole deviations using X-ray observations \cite{Bambi_2011_1, Bambi_2011_2}.

\subsection{Johannson-Psaltis spacetime}

The JP metric \cite{Psaltis_2011} is another example of a spacetime that describes parametric deviations from Kerr. This metric originates with the familiar Kerr metric, to which deviations are added that are  proportional to
\begin{equation}
    h(r, \theta) = \sum_{k=0}^\infty \left(\epsilon_{2k} + \epsilon_{2k+1}\frac{Mr}{\Sigma} \right)\left( \frac{M^2}{\Sigma}\right)^k
\end{equation}
where $\Sigma=r^2 + a^2 \cos^2 \theta$ and $\epsilon_k$ are deviation parameters. It was shown in \cite{Psaltis_2011} that asymptotic flatness requires $\epsilon_0 = \epsilon_1 = 0$. 
In the following, we will consider the special case of $\epsilon_k = 0$ for $k>3$. In that case, the deviations from Kerr are proportional to the ``quadrupole'' term
\begin{equation}
    h(r, \theta) = \epsilon_3 \frac{M^3 r}{\Sigma^2}
\end{equation}
and the resulting metric is of the form
\begin{widetext}
\begin{eqnarray}\label{eq:transfer_main}
    ds^2 =&& -\left(1+h(r,\theta)\right)\left(1-\frac{2Mr}{\Sigma}\right)dt^2-\frac{4aMr\sin^2\theta}{\Sigma}\left(1+h(r,\theta)\right)dt d\phi + \frac{\Sigma\left(1+h(r,\theta)\right)}{\Delta + h(r,\theta) a^2 \sin^2\theta}dr^2\nonumber \\
    &&+\Sigma d\theta^2 + \left\{\sin^2\theta \left(r^2 + a^2 +\frac{2 a^2 Mr \sin^2\theta}{\Sigma}\right) + h(r,\theta) \frac{a^2\left(\Sigma + 2Mr\right)\sin^4\theta}{\Sigma}\right\}d\phi^2\\\nonumber
\end{eqnarray}
\end{widetext}
where $\Delta = r^2 - 2Mr+ a^2$. In the limit $\epsilon_3 \rightarrow 0$, the JP metric reduces to Kerr. From the asymptotic structure of the metric, it is clear that this JP metric has a quadrupole deviation from Kerr proportional to $\epsilon_3 M^3$ in the Newtonian limit.

As in the case of the MN metric, the properties of the event horizon in the JP metric are quite different from  Kerr. For negative values of $\epsilon_3$, the event horizon is always closed. However, for positive values of $\epsilon_3$, for any given value of the spin $a$ there exists a maximum value of $\epsilon_3$ beyond which the event horizon is no longer closed \cite{Psaltis_2011}. A `break' in the event horizon appears on the equatorial plane, transforming the central object into a naked singularity. In our analysis, we avoid this kind of pathology by excluding the part of the parameter space that corresponds to the aforementioned singularity. 

Since its introduction, the JP metric has seen widespread use in testing GR under various frameworks. The simplicity of the parametrized metric makes it a particularly interesting tool in performing null tests of gravity. It has been used in conjunction with X-ray observations to study accretion disks~\cite{Kong:2014wha, Bambi:2015ldr}. In the context of the GW  ringdown, this metric has been used in \cite{Glampedakis_2017, Carson:2020iik}, where the analysis is performed assuming small $\epsilon_3$. Typical values of $\epsilon_3$ that agree with data  may be as large as $\sim 14$, which casts some doubt on the validity of such an assumption. We avoid these restrictions in our study, and allow $\epsilon_3$ to vary within a wide range while estimating the posterior distribution from data.

\section{Eikonal Approximation}\label{sec:eikonal}

Given an alternative theory of gravity, the first step toward testing it with  ringdown signals is to devise a way to calculate its QNMs. This is not easy to do in arbitrary spacetimes. The usual prescription involves solving the BH perturbation equations under appropriate boundary conditions. On the
Kerr spacetime,
such equations can be solved by separation of variables, but this is typically not possible on more generic background and/or beyond GR.
This calls  for  alternative formalisms/approximations for calculating QNMs.

Fortunately, QNMs and  unstable null geodesics of the background are closely related in the the eikonal, or short-wavelength, approximation.
The latter
 was initially investigated by Press \cite{Press_1971} (see also Goebel's comment on the same \cite{Goebel_1972}). It has since been understood that the real part of the QNM frequencies in Kerr is related to the frequency of the unstable light ring, and their imaginary part is related to its instability timescale. Ferrari and Mashhoon provided seminal contributions to extending this line of thought \cite{Ferrari_1984, Mashhoon_1985}. In more recent times, this has also been explored in \cite{Dolan_2010, Yang_2012, Yang_2013_1, Yang_2013_2, Cardoso:2008bp}.

QNMs are indexed by three numbers - $n$, $\ell$ and $m$. $n$ is the overtone number and $\ell$ and $m$ are the indices of the $\left(\ell,m\right)$ multipole.
In the eikonal regime ($\ell,m\gg1$), the Kerr QNM frequency for the maximally corotating mode $\ell=m$ is
\begin{equation}\label{eq:eikonal}
    \omega_{\text{QNM}} = \ell \Omega - i \gamma\left(n+\frac{1}{2} \right)\,,
\end{equation}
where $\Omega$ is the orbital frequency of light rays on the unstable equatorial circular orbit, and $\gamma$ is the Lyapunov exponent of the same orbit.
The  Lyapunov exponent
 characterizes the timescale on which
 the cross section of a congruence of null rays
 increases
  under radial perturbations. Though Eq. \eqref{eq:eikonal} is only strictly valid  for $\ell,m\gg1$, it works remarkably well for low values of $\ell$ as well \cite{Berti:2005eb, Iyer:1986nq}.

The eikonal approximation is clearly an incredibly powerful tool, because one only needs to calculate the radius and Lyapunov exponent of the unstable circular equatorial null orbit to calculate the QNM frequencies, without having 
to solve the BH pertubation equations.
In the following, as in Ref.~\cite{Glampedakis_2017}, we will make the physically reasonable 
(although unproven\footnote{Although the correspondence between QNM frequencies and unstable circular orbits is not necessarily automatic beyond GR, it can be proven in some regimes. For instance, if one makes the assumption that the equation for the gravitational perturbations has principal part given by 
$\Box h_{\mu\nu}$, where  the $\Box$ operator is computed with the modified (e.g. 
MN or JP) metric, then the correspondence can be established rather rigorously. See, for e.g., discussion in Sec. IIB of \cite{Volkel:2020daa} and in Sec. III of 
\cite{Ghosh:2023etd}.}) 
assumption that the relation between null orbits and QNM frequencies also holds on  (stationary and axisymmetric) backgrounds different from Kerr  and in theories beyond GR (see however~\cite{Khanna:2016yow}).

For equatorial null geodesics in a generic stationary and axisymmetric spacetime, one has
\begin{equation}\label{eq:null-geo}
    g_{tt}\left(u^t\right)^2 + 2g_{t\varphi}u^t u^\varphi + g_{rr}\left(u^r\right)^2 + g_{\varphi\varphi}\left(u^\varphi\right)^2 = 0\,,
\end{equation}
where $(u^t, u^r, u^\theta, u^\varphi)$ is the four-velocity. From the spacetime symmetries, the conserved quantities are the energy $E$ and angular momentum $L$, defined by $E=-u_t$ and $L=u_\varphi$. Accordingly, one has
\begin{equation}\label{eq:ut}
    u^t = \frac{1}{g_{t\varphi}^2-g_{tt}g_{\varphi\varphi}} \left(g_{\varphi\varphi}E+g_{t\varphi}L\right)\,,
\end{equation}
and
\begin{equation}\label{eq:uphi}
    u^\varphi = -\frac{1}{g_{t\varphi}^2-g_{tt}g_{\varphi\varphi}}\left(g_{tt}L + g_{t\varphi}E\right)
\end{equation}
Using then Eqs.~\eqref{eq:ut} and \eqref{eq:uphi} in Eq.~\eqref{eq:null-geo}, one obtains
\begin{align}\label{eq:ur}
    \left(u^r\right)^2 &= \frac{1}{g_{rr}\left(g_{t\varphi}^2-g_{tt}g_{\varphi\varphi}\right)} \left(g_{\varphi\varphi} E^2  + 2g_{t\varphi}EL + g_{tt}L^2\right)\nonumber,\\
    &\equiv V_{\text{eff}}
\end{align}
This equation is of the form $(u^r)^2= V_\text{{eff}}$, where $V_{\text{eff}}$ is an effective potential for the radial motion. The turning points of the orbital motion then correspond to locations at which $u^r$ vanishes. In other words, the radial coordinate $r_0$ at the turning points must satisfy
\begin{equation}\label{eq:tp1}
    g_{\varphi\varphi}(r_0) + 2g_{t\varphi}(r_0) b + g_{tt}(r_0) b^2 = 0\,,
\end{equation}
where $b=L/E$ is the orbit's impact parameter. The angular frequency can then be defined from Eqs. \eqref{eq:ut} and \eqref{eq:uphi} as $\Omega = u^\varphi/u^t$. Using this in Eq.~\eqref{eq:null-geo},
one has
\begin{equation}\label{eq:tp2}
    g_{tt}(r_0) + 2 g_{t\varphi}(r_0)\Omega_0 + g_{\varphi\varphi}(r_0)\Omega_0^2 = 0\,,
\end{equation}
where $\Omega_0 \equiv \Omega(r_0)$. From Eqs.~\eqref{eq:tp1} and \eqref{eq:tp2}, one  finds that at the turning points
one has
\begin{equation}\label{eq:omega_b_relation}
    \Omega_0 = 1/b\,.
\end{equation}
For circular orbits to exist at $r=r_0$, 
$r_0$ must also extremize the effective potential, i.e.
$r_0$ must satisfy
 the  conditions
\begin{equation}\label{eq:light-ring}
    V_{\text{eff}}(r_0)=0 \quad \quad V_{\text{eff}}'(r_0)=0\,.
\end{equation}
Solving these equations, one obtains the radius of the unstable light ring $r_{\text{ph}}=r_0$ and the impact parameter $b$. 

Next, let us turn our attention to  calculating the Lyapunov exponent, which is related to the decay time of the QNMs. We start with a simple change of radial coordinate, $\mathcal{R}=1/r$. Eliminating $u^r$ and $u^\varphi$ from Eqs.~\eqref{eq:uphi} and~\eqref{eq:ur}, one obtains
\begin{eqnarray}
    \left(\frac{d\mathcal{R}}{d\varphi}\right)^2 &&= \mathcal{R}^4 \frac{\left(g_{t\varphi}^2 - g_{tt}g_{\varphi \varphi}\right)^2}{\left(g_{tt} b + g_{t\varphi}\right)^2}V_\text{eff} \nonumber\\
    &&\equiv f(\mathcal{R})\,.\label{lyap}
\end{eqnarray}

For a circular orbit of radius $r_0$, or equivalently $\mathcal{R}_0=1/r_0$, $f(\mathcal{R}) = f'(\mathcal{R}) = 0$. Perturbing the circular null orbit, one can then write
\begin{equation}\label{eq:perturbexpansion}
    \mathcal{R} = \mathcal{R}_0 + \epsilon \mathcal{R}_1 + O(\epsilon^2)    
\end{equation}
where $\epsilon\ll1$ is a perturbative parameter. It can then be shown \cite{Glampedakis_2017} that Eq.~\eqref{lyap} leads to 
\begin{equation}
    \frac{d\mathcal{R}_1}{d\varphi} = \pm \kappa_0 \mathcal{R}_1
\end{equation}
where 
\begin{align}
    \kappa_0^2 = \frac{1}{2\mathcal{R}_0^4}f''(\mathcal{R}_0)
    =\frac{1}{2}\frac{V''_{\text{eff}}}{\left(u^\varphi\right)^2}
\end{align}
is to be evaluated at $r=r_0$.
This leads to solutions for $\mathcal{R}_1$ of the form $\mathcal{R}_1 = \mathcal{A}e^{\pm \kappa_0 \varphi}$, where $\mathcal{A}$ is a constant. With $\Omega_0$ known, one can write $\varphi = \Omega_0 t + \varphi_0$ where $\varphi_0$ is a constant. Using this in Eq.~\eqref{eq:perturbexpansion},
one arrives at
\begin{equation}
    \mathcal{R} = \mathcal{R}_0 + \epsilon \mathcal{A} e^{\pm \gamma_0 t}\,,
\end{equation}
where 
\begin{align}\label{eq:lyapunov}
    \gamma &= \kappa_0 \Omega_0 \nonumber,\\
    &=\sqrt{\frac{1}{2}\frac{V''_{\text{eff}}}{\left(u^t\right)^2}}
\end{align}

To summarize, $r_0$ and $\Omega$ can be found by imposing circular orbit conditions on the effective potential. Subsequently, $\gamma$ can be calculated at $r_0$, providing us with all the tools necessary to calculate QNM frequencies with Eq.~\eqref{eq:eikonal}. Although this process is quite straightforward, solving the light ring equations can be non-trivial and analytical solutions may not always be possible, in which case one has to resort to numerical solutions. 

\section{Methodology}\label{sec:methodology}

The GW  signal from the ringdown of a binary BH merger can be decomposed into a superposition of damped sines and cosines. If $h_+$ and $h_\times$ are the two GW polarisations, 
\begin{equation}\label{eq:ringw_def}
    h_+ - i h_\times = \sum_{\ell m} h_{\ell m}\left(t\right)_{-2}Y_{\ell m}\left(\iota,\varphi\right)\,,
\end{equation}
where $_{-2}Y_{\ell m}$ are the spin-weighted spherical harmonics.\footnote{The $_{-2}Y_{\ell m}$ are used as an approximation to the spin-weighted spheroidal harmonics $_{-2}S_{\ell m}$, which reduce to $_{-2}Y_{\ell m}$ when $a\tilde{\omega}_{lm}=0$, where $\tilde{\omega}_{lm}$ is the QNM frequency and $a$ is the BH spin. It was however shown that $_{-2}S_{\ell m} \approx _{-2}Y_{\ell m}$ also for non-zero spins~\cite{Berti:2005gp}.} 

The functions $h_{\ell m}\left(t\right)$ can be expressed as
\begin{equation}\label{eq:RD-model}
    h_{\ell m}\left(t\right) = \sum_n A_{\ell mn} e^{-t/\tau_{\ell mn}} e^{-i\omega_{\ell mn}t + \phi_{\ell mn}}\,.
\end{equation}
Here, $A_{\ell mn}$ and $\phi_{\ell mn}$ are the amplitude and phase of the $\left(\ell,m,n\right)$ mode, and $\omega_{\ell mn}$ and $\tau_{\ell mn}$ are its frequency and damping time respectively.

For quasi-circular binary mergers,
the dominant mode is $\ell=m=2$, and in our analysis of GW150914 we restrict ourselves to it. For fixed $(\ell,m)$, the QNMs can be arranged by their overtone number $n$, with $n=0$ being the `fundamental' mode and  higher values denoting the `overtones'. The question of the detection of overtones in the GW150914 signal is still hotly debated. Although \cite{Isi:2019aib, Isi:2021iql} claimed a positive detection of the $n=1$ overtone of the $\ell=m=2$ mode, subsequent work~\cite{Cotesta:2022pci} questioned this claim.  In the absence of a consensus on this topic, we shall  restrict our analysis of GW150914 to the fundamental mode alone.

The ringdown parameter estimation analysis of GW150914 is obviously accompanied/preceded by the full signal's analysis, which includes the inspiral and merger phases as well. As a result, the sky position, time of merger, and luminosity distance are known quite well. These can then be considered as fixed parameters in the ringdown analysis. The detector measures $h_{\text{strain}} = F_+ h_+ + F_\times h_\times$, where $F_+$ and $F_\times$ are the detector pattern functions, which depend on the right ascension (=1.95 rad), declination (=-1.27 rad), polarization (=0.82 rad). Following previous studies on the GW150914 ringdown \cite{Isi:2019aib, Cotesta:2022pci}, we also fix ($\iota$, $\varphi$)=($\pi$,0). For the likelihood computation, we use the time domain method  described in \cite{Isi:2021iql}. The latter consists of truncating the signal just before the ringdown phase and using the noise covariance matrix to calculate the likelihood in the  time domain, without venturing into the frequency domain. This avoids potential data loss due to windowing during fast Fourier transforms. We use the software package \texttt{ringdown} \cite{maximiliano_isi_2021_5094068} (based on \cite{Isi:2021iql}) for data truncation, calculation of the noise covariance matrix and  detector pattern functions. We use the publicly available sampler \texttt{dynesty} \cite{dynesty_2022_6609296}, which uses a nested sampling algorithm,  to sample the parameter space and calculate the posterior distributions. 
\begin{figure*}
    \centering
    \includegraphics[width=0.8\textwidth]{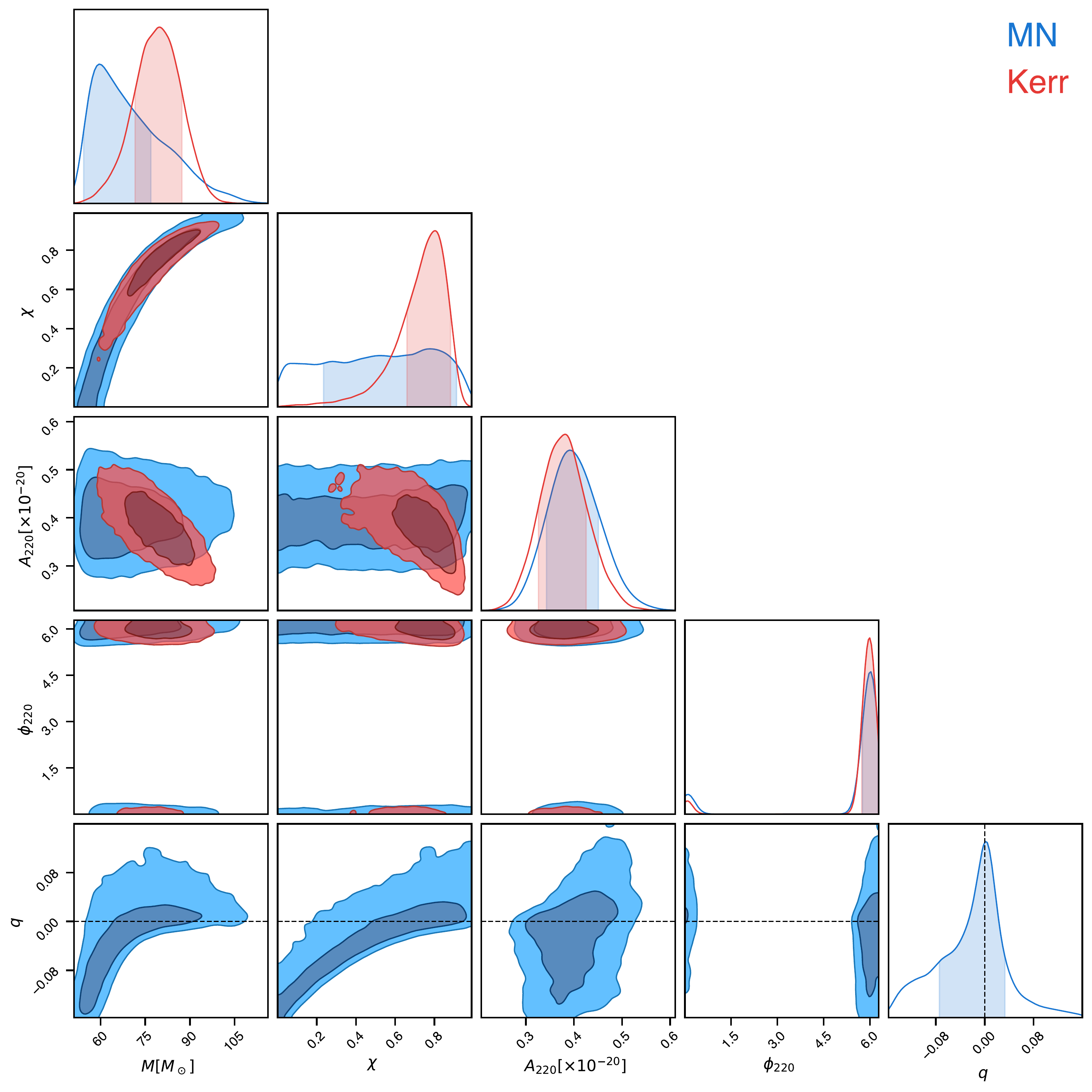}
    \caption{Posterior distributions for the parameters obtained from the GW150914 data,  assuming MN and Kerr background spacetimes. The darker shade in the contours denotes the 1$\sigma$ confidence level, while the lighter shade denotes the 2$\sigma$ level. The shaded areas in the one-dimensional histograms correspond to  1$\sigma$ confidence levels. $q=0$ (where the MN metric reduces to Kerr) is also highlighted in the distribution with a dotted black line.}
    \label{fig:MNpost}
\end{figure*}
As for the waveform model, we use the procedure outlined in section \ref{sec:eikonal} to calculate the QNMs for both metric ansatze under consideration. We solve the light ring equations numerically to obtain QNM frequencies as functions of mass, spin and  deviation parameter ($q$ for the MN metric and $\epsilon_3$ for the JP metric). The resulting GW strain is then calculated by projecting $h_+$ and $h_\times$ onto the detector using the  pattern functions. We sample the following parameters - remnant mass $(M)$, dimensionless spin $(\chi)$,   amplitude $(A_{\ell mn})$ and phase $(\phi_{\ell mn})$ of the ($\ell, m, n$) mode, along with the  deviation parameter ($q$ or $\epsilon_3$). In our analysis of GW150914, we consider only the fundamental $\ell=m=2$ mode. This means that our model has five free parameters, including $q$ or $\epsilon_3$.
Flat priors are assumed on all parameters - $M\in [20,200]M_\odot$, $\chi \in [0,0.99]$, $A_{220} \in [0,5\times10^{-20}]$ and $\phi_{220} \in [0, 2\pi]$. As for the  deviation parameters, we choose $q \in [-0.16, 0.16]$ and $\epsilon_3 \in [-30,100]$.  As for the data itself, we consider the GW150914 data from both the Hanford and Livingston detectors. We choose the truncation time for the signal from these detectors to be respectively 1126259462.423 s and 1126259462.4160156 s, and a sampling rate of 4096 Hz. The posteriors obtained are presented in Sec \ref{sec:results}.

\section{Results}\label{sec:results}

Let us start by applying the formalism described above to the MN geometry.
As is clear from Eqs. \eqref{eq:MN1} - \eqref{eq:MN3}, the MN metric is quite involved to work with, and developing BH perturbation theory in such a spacetime is a non-trivial task. In fact, the perturbation equations in general do not separate. This is also the case for many alternative theories of gravity. Fortunately, the procedure outlined in Sec. \ref{sec:eikonal} is still within reach, and Eq. \eqref{eq:light-ring} can indeed be solved.

Starting from the equatorial MN metric, we can easily calculate the effective potential $V_{\text{eff}}$ and its radial derivative. We then solve  the system \eqref{eq:light-ring} numerically to obtain  $r_{\text{ph}}$ and $b$ as functions of the spin $\chi$ and the quadrupole parameter $q$. The remnant mass enters as a prefactor. We use the data for GW150914, conditioned as described in Sec. \ref{sec:methodology}. The eikonal QNM frequencies are used along with the data to calculate the likelihood for arbitrary values of the sampled parameters - $(M, \chi, A_{220}, \phi_{220}, q)$. As mentioned above, we use \texttt{dynesty} and the resulting posterior distributions are shown in Fig. \ref{fig:MNpost}.
\begin{figure*}
    \centering
    \includegraphics[width=0.8\textwidth]{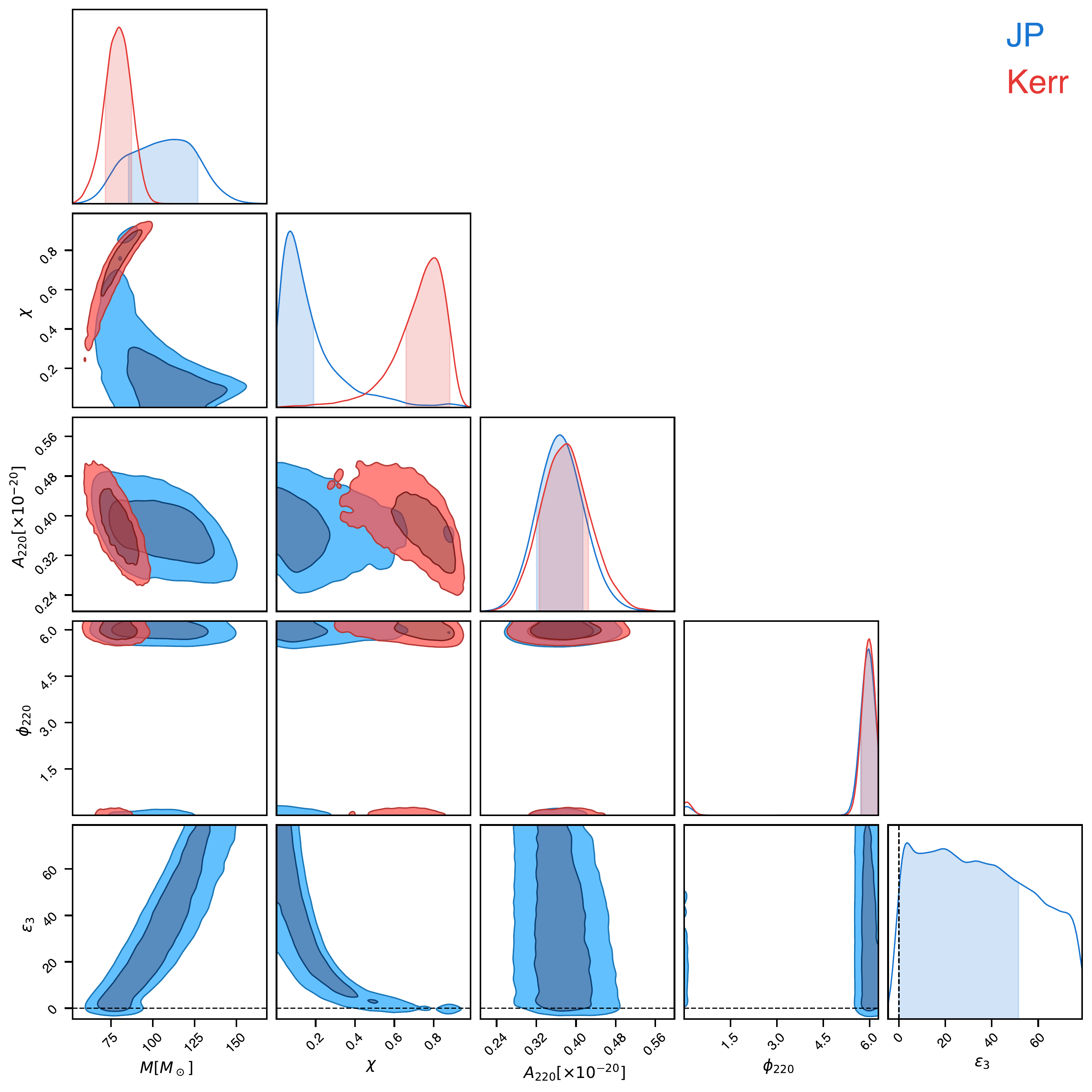}
    \caption{The same as in Fig.~\ref{fig:MNpost}, but for the JP spacetime (with large priors on the remnant mass). $\epsilon_3=0$ (where the JP metric reduces to Kerr) is also highlighted in the distribution with a dotted black line.}
    \label{fig:JPpost}
\end{figure*}
As can be clearly seen, the posterior distribution for $q$ is strongly peaked around zero, which is the Kerr case. This suggests that the Kerr hypothesis is indeed favoured by the data, although $|q|$ as large as $\sim 0.05$ are also compatible with the data at 95\% confidence level. Another point to note is that the marginalized bounds on $\chi$ are uninformative. This is because a change in $q$ can compensate for variations introduced by changes in $\chi$. This is in stark contrast to the Kerr metric,
where this degeneracy obviously does not exist and the spin can be bound quite well. It can also be seen that the recovered posterior for $M$ is different from its Kerr counterpart, again due to the degeneracy with $q$. 


The JP metric has been previously used for ringdown analyses \cite{Carson:2020iik}, but using the Fisher matrix formalism. Moreover, the QNMs were calculated using the prescription described in \cite{Glampedakis_2017}, which involves an expansion in the `small' deviation parameter $\epsilon_3$. However, it is unclear whether this expansion is justified by the data, which may not bound  $\epsilon_3$ to a significant degree. In this paper, we use the full form of the JP metric, without imposing any restrictions on the value of $\epsilon_3$, and adopt  a Bayesian framework in place of a Fisher matrix one for the estimation of the posterior distributions.
\begin{figure*}
    \centering
    \includegraphics[width=0.8\textwidth]{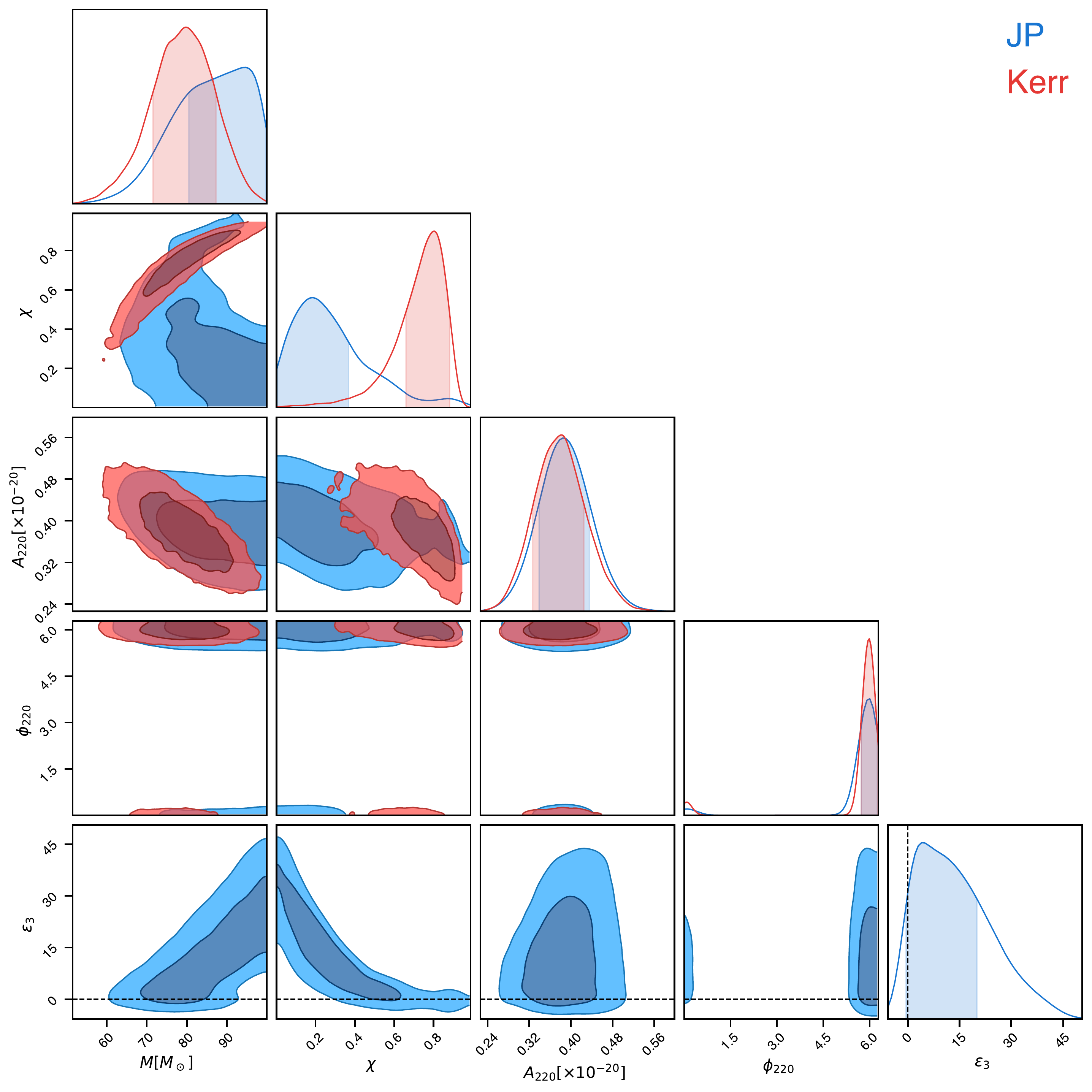}
    \caption{The same as in Fig.~\ref{fig:JPpost}, but with a remnant mass prior $M\in [55,100]M_\odot$.}
    \label{fig:JPpostIMR}
\end{figure*}
Similar to the MN metric case, we use the equatorial JP metric and calculate the QNM frequencies following the procedure outlined in Sec. \ref{sec:methodology}. 
The estimated posterior distributions are shown in Fig. \ref{fig:JPpost}.

While the distribution for $\epsilon_3$ does have a peak near the Kerr value $\epsilon_3=0$, it is highly skewed towards $\epsilon_3>0$, with virtually no support for  negative value. The estimated posteriors for mass $M$ are also quite different from the Kerr and MN ones. In particular, they show support for comparatively larger values of $M$ and smaller values of $\chi$ than those predicted for Kerr metric~\cite{Isi:2019aib}.

The differences between the posteriors for Kerr and JP are quite striking, especially for $\chi$. This can be understood by examining the analytic form of the QNM frequencies of the JP metric. While such analytic expressions are difficult to compute for  arbitrary $\epsilon_3$ (we calculate them by solving Eq~\eqref{eq:light-ring} numerically), they can be obtained for small values of $\epsilon_3$. Ref.~\cite{Carson:2020iik} found that for small $\epsilon_3$
\begin{align}\label{eq:JP-post-kerr}
    \omega_\text{R}^\text{JP} &= \omega_\text{R}^\text{K} + \epsilon_3 \left(\frac{1}{81\sqrt{3}M} + \frac{10}{729M}\chi + \frac{47}{1458\sqrt{3}M}\chi^2\right),\\
    \omega_\text{I}^\text{JP} &= \omega_\text{I}^\text{K} - \epsilon_3 \left(\frac{1}{486M}\chi + \frac{16}{2187\sqrt{3}M}\chi^2\right)
\end{align}
According to Eq.~\eqref{eq:JP-post-kerr}, the effect of the deviation parameter can be countered by smaller values of $\chi$ and larger values of $M$. This is reflected in the posterior distribution for the JP metric, which favours higher $M$ and lower $\chi$ compared to  Kerr.

Note however that the mass of the remnant can be estimated independently from the ringdown and from the inspiral phase. These measurements must of course be consistent~\cite{Ghosh:2016qgn}. Unfortunately, estimating the mass from the inspiral requires knowledge of the (non-GR) field equations, which do not simply follow from knowledge of the JP geometry. Therefore, performing such a consistency test is not possible in our setting. We can, however, assume that deviations from Kerr and GR are
``small'', and that the inspiral determination of the mass within GR is approximately correct even if the remnant is described by the JP spacetime (rather than by the Kerr one).
 Under this assumption, we impose a conservative prior bound on the remnant mass $M\in [55,100]M_\odot$. (Note that the bound from the inspiral calculated using GR is actually much tighter \cite{LIGOScientific:2016vlm}. We deliberately choose a larger bound to account for the 
different remnant geometry.) The resulting posteriors for the JP metric  are shown in Fig. \ref{fig:JPpostIMR}. While the posteriors for $\epsilon_3$ are still skewed towards  positive values, they are now much smaller and  consistent with those from \cite{Carson:2020iik, Kong:2014wha}. 

\begin{figure}
    \centering
    \includegraphics[width=0.95\linewidth]{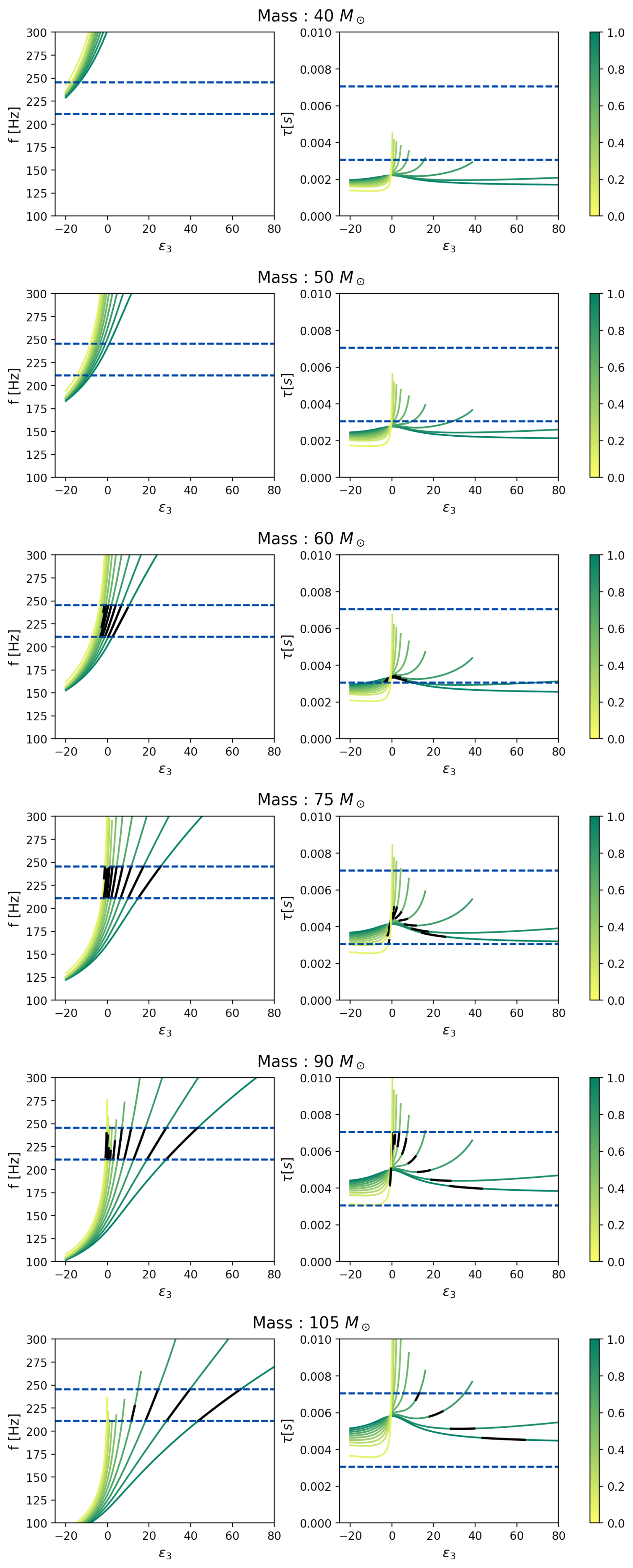}
    \caption{Dependence of the frequency and damping time of the dominant 220 mode on $\epsilon_3$, for different values of spin and remnant mass. Different spins are denoted by the color axis, which goes from 0 to 1. The blue lines indicate the GW150914 bounds on frequency and damping time. The black lines denote the parameter values for which both bounds are satisfied.}
    \label{fig:fr-tau-JP}
\end{figure}

We now turn our attention to understanding the asymmetry of the posterior distribution of $\epsilon_3$ about zero. A linear ringdown model of the form of Eq.~\eqref{eq:RD-model} implies that while performing parameter estimation, the sampler is trying to fit the data to some unknown frequency and damping time along with  an amplitude and a phase. The frequency and damping time depend on the remnant mass, spin and  deviation parameter. For the GW150914 data, the fitted frequency and damping time lie in the range $\sim$210-240 Hz and $\sim$3-7 ms, respectively. The sampler then searches the parameter space for combinations of mass, spin and  deviation parameter that correspond to values in this range. In Fig. \ref{fig:fr-tau-JP}, we show the dependence of QNM frequency and damping time on spin and $\epsilon_3$, for different values of the remnant mass. The solid black lines denote the portion of the parameter space where the bounds on frequency and damping time are simultaneously satisfied. These lines are not present for  low values values of $M$ ($\lesssim 50 M_\odot$), because the QNM frequencies are too large and the damping times are too small to agree with the data. As $M$ increases, we start to observe that some points in the parameter space become accessible. For intermediate values of $M<60 M_\odot$, the predicted frequencies lie within the observed bounds for both positive and negative values of $\epsilon_3$, but the bounds on the damping time are satisfied only for positive values (the support for negative values values is minimal, as seen in Fig. \ref{fig:fr-tau-JP}). For higher values of $M>60 M_\odot$, the opposite is true, i.e. the bounds on the damping time are satisfied by a wide range of values of $\epsilon_3$, but the frequency bounds are met mostly for $\epsilon_3>0$, with minimal support for negative values. In other words, over a wide range of $M$, $\epsilon_3 >0$ leads to frequencies and damping time that simultaneously lie within their respective bounds. This results in $\epsilon_3>0$ being favoured in the posterior distribution shown in Figs.~\ref{fig:JPpost} and \ref{fig:JPpostIMR}.

With the asymmetry in the posterior distribution of $\epsilon_3$ now understood, we turn our attention to the difference in the estimates of $q$ and $\epsilon_3$. Upon a cursory glance, both denote a quadrupole deviation from Kerr. Multipole moments can be calculated for axisymmetric, asymptotically flat and vacuum spacetimes by performing an asymptotic expansion of the metric functions \cite{Ryan:1996nk}. This method can be applied to the MN metric and the quadrupole moment $Q_{\text{MN}}$ is $Q_{\text{Kerr}} - q M^3$. For the JP metric, using a similar asymptotic expansion one finds the quadrupole moment to be $Q_{\text{JP}} = Q_{\text{Kerr}} + \epsilon_3 M^3$ \cite{Glampedakis_2017}. The JP metric is, however, not a vacuum spacetime if it is to satisfy the Einstein field equations, which casts doubts on the above result. 

Even after restricting the mass priors (in Fig. \ref{fig:JPpostIMR}), one can still observe an order of magnitude difference in the posterior bounds on $q$ and $\epsilon_3$. In the eikonal prescription used in this paper, the central quantity of interest is the effective potential $V_{\text{eff}}$ defined in Eq. \eqref{eq:light-ring}. Let us then turn to this potential to understand this discrepancy. Since the MN potential is too cumbersome to work with, we perform an expansion assuming small spin and small quadrupolar deviation ($q$ or $\epsilon_3$).

Solving then Eq.~\eqref{eq:light-ring} at leading order, the impact parameter is
\begin{align}
    b_{\text{ph}}^{\text{MN}}/M &\approx 3\sqrt{3} - 2\chi + q\,\\
    b_{\text{ph}}^{\text{JP}}/M &\approx 3\sqrt{3} - 2\chi - \frac{\epsilon_3}{6\sqrt{3}}
\end{align}
in the two geometries. The impact parameter is related to the real part of the QNM frequency through Eqs. \eqref{eq:eikonal} and \eqref{eq:omega_b_relation}.
For the two geometries to predict the same QNM frequencies, we must have
\begin{equation}\label{eq:order-diff}
    \epsilon_3 = -6\sqrt{3}q \approx -10 q\,
\end{equation}

This relation, however, is not sufficient to explain the posteriors of $q$ and $\epsilon_3$, which differ by a factor $\sim O(100)$. To properly understand this discrepancy, and whether a mapping between the two metrics is even possible at all, we also look at the imaginary part of the QNM frequencies, as given by Eqs. \eqref{eq:lyapunov} and~\eqref{eq:eikonal}, under the same small parameter expansion. We find
\begin{align}
    \gamma^{\text{MN}} &\approx \frac{1}{3\sqrt{3}}+\frac{1}{54}\left(-13+12\sqrt{3}\right)q,\\
    \gamma^{\text{JP}} &\approx \frac{1}{3\sqrt{3}}\,.
\end{align}
This shows that at the lowest order in spin and deviation parameter, $\gamma^{\text{JP}}$ is independent of $\epsilon_3$ whereas $\gamma^{\text{MN}}$ depends on $q$. Therefore, an exact mapping between the parameters $q$ and $\epsilon_3$ does not exist, which explains the different posteriors for the two parameters.

We also calculate $\log_{10} \mathcal{B}_{\text{Kerr}}^{\text{MN}}$ and $\log_{10} \mathcal{B}_{\text{Kerr}}^{\text{JP}}$ which are the $\log_{10}$-Bayes factors for the MN and JP metrics compared to Kerr. The value of these factors  indicates  whether the MN and/or JP hypothesis fit the data better than Kerr. We find $\log_{10} \mathcal{B}_{\text{Kerr}}^{\text{MN}} = -0.71$ and $\log_{10} \mathcal{B}_{\text{Kerr}}^{\text{JP}} = -0.25$. The negative values indicate that the Kerr hypothesis is better supported by the GW150914 data.

\section{Future Detectors}\label{sec:future-detectors}

\begin{figure}
    \centering
    \includegraphics[width=\linewidth]{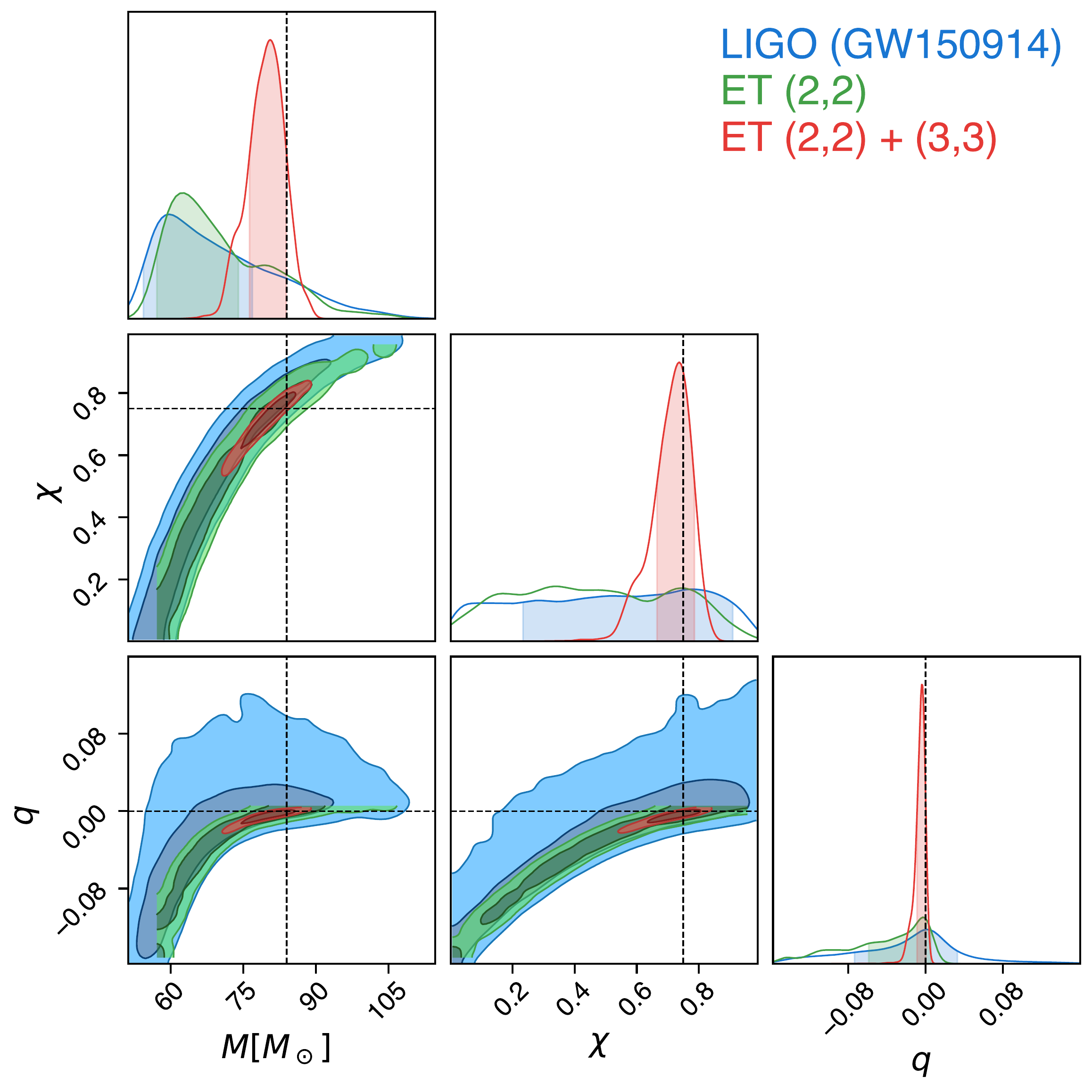}
    \caption{Posterior distributions
    for $M$, $\chi$ and $q$ obtained using ET. In red, we show the distributions observed while including the (3,3) mode in addition to the dominant (2,2) mode. The distributions with only the (2,2) mode are shown in green. For reference, the posteriors for GW150914 (which have similar remnant mass and spin) observed with LIGO are shown in blue.}
    \label{fig:MN-ET}
\end{figure}

\begin{figure}
    \centering
    \includegraphics[width=\linewidth]{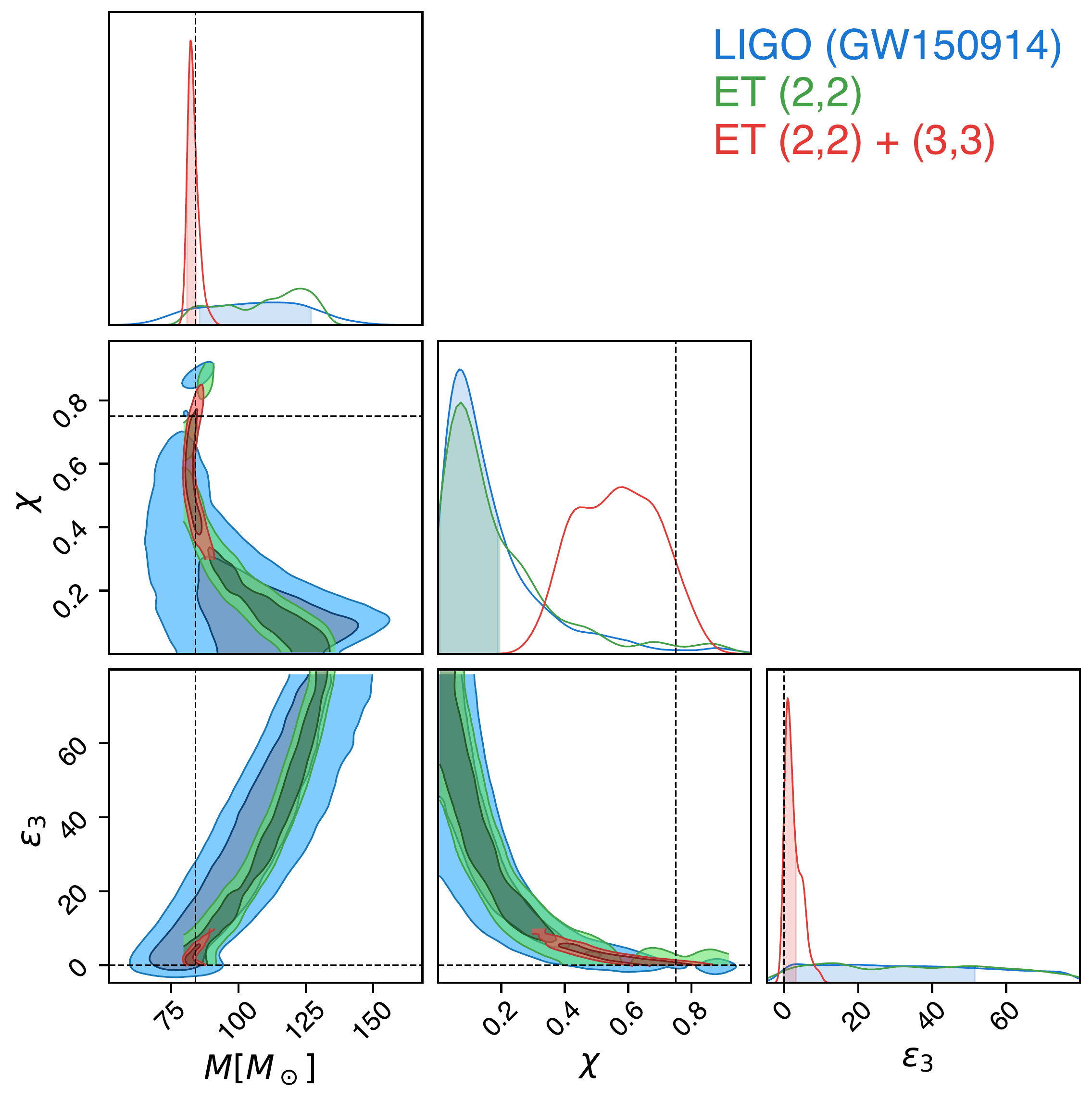}
    \caption{The same as in Fig.~\ref{fig:MN-ET}, but for the JP case.}
    \label{fig:JP-ET}
\end{figure}

The LIGO-Virgo-Kagra (LVK) detectors, along with LIGO-India in the coming years, make up the second generation of interferometric GW detectors. These will be followed up by  third generation detectors - the Einstein Telescope (ET) \cite{Punturo:2010zz} and Cosmic Explorer (CE) \cite{Reitze:2019iox} - and by space based detectors like  LISA \cite{LISA:2017pwj}. These future detectors are designed to allow for lower noise levels, boosting the SNR of detected sources in comparison to LVK. High SNR is an important requirement of ringdown analysis because of the fast decay of QNMs. Indeed low SNRs prevent the use of most GW events for ringdown analyses. GW150914 (and GW190521 \cite{LIGOScientific:2020iuh,Capano:2021etf}) is an outlier in this regard.  Future detectors are expected to improve on this by providing more eligible sources and higher SNRs~\cite{Berti:2016lat}. Additionally, higher SNRs will likely lead to detection of higher modes. These modes can also potentially break the degeneracy in $M$ and $\chi$ that is observed due to the inclusion of quadrupole deviations in both the MN and JP geometries.

Higher SNR is beneficial for parameter estimation and will lead to an overall shrinkage in the posterior volume. We simulate two different types of signals detectable by ET - one injection containing only the $\ell=m=2$ mode and one injection containing the $\ell=m=3$ in addition to the dominant $\ell=m=2$ mode. The remnant mass and spin of the injected signals are $M=84 M_\odot$ and $\chi=0.75$, which are similar to GW150914, allowing for a comparison between the posterior bounds across detectors. In order to have a non-zero contribution from the $\ell=m=3$ mode, we assume $\left(\iota, \varphi \right) = (\pi/3,0)$ (in Eq.~\ref{eq:ringw_def}) for the injections under consideration. As for the amplitude and phase of the higher mode, we use the fitting formulae provided in \cite{Forteza:2022tgq}. With the $\ell=m=2$ mode as the reference, the amplitude and phase of the higher mode are redefined as
\begin{equation}
    A_{330}^R = \frac{A_{330}}{A_{220}}, \qquad \delta\phi_{330} = \frac{3}{2}\phi_{220} - \phi_{330}
\end{equation}

We consider a system where the mass ratio is 3. Using the fitting formulae from \cite{Forteza:2022tgq}, we get $ A_{330}^R = 0.255$ and $\delta\phi_{330} = 2.95$. The parameters $A_{220}$ and $\phi_{220}$ are randomly chosen to be $0.44\times 10^{-20}$ and $5.37$. Moreover, we inject a Kerr signal, with no quadrupole deviations. To this injection, we add simulated noise  compatible with the design sensitivity of ET (ET-D sensitivity curve), considering for simplicity a single interferometer in the planned ET ``xylophone'' configuration~\cite{Hild:2010id}. The SNRs of the resulting signals are 176 for the $\ell=m=2$ only mode, and 179 with the inclusion of the $\ell=m=3$ mode (as a comparison, the ringdown SNR for GW150914 is $\sim$13). While sampling, we use the parameters $\left(M, \chi, A_{220}, A_{330}^R, \phi_{220}, \phi_{330} \right)$, along with the deviation $q/\epsilon_3$ for the MN/JP geometries.

The simulated signal is analyzed according to the process outlined in Sec. \ref{sec:methodology}, with the LIGO detector swapped for ET. We perform this analysis both in the presence and absence of the higher mode for the MN and JP geometries. The results are shown in Figs. \ref{fig:MN-ET} and \ref{fig:JP-ET}, and compared to those obtained with LIGO. The posterior distributions obtainead from LIGO GW150914 (in blue) and those from ET (2,2) (in green) are mostly similar, but with a general reduction in the posterior volume. The lower noise levels imply that the QNM frequencies and  damping times can be estimated with  greater accuracy, but that does not eliminate the degeneracies among $M$, $\chi$ and $q$ ($\epsilon_3$). The decrease in posterior volume in the contours is, on the other hand, quite visible. As for the quadrupolar deviation, the $1\sigma$ limits of the posterior distribution decreases by $\sim 24\%$ for $q$, while remaining mostly unchanged for $\epsilon_3$, compared to those from LIGO.

Including the $\ell=m=3$ mode, however, has a much more drastic effect on the posterior distribution. The degeneracy in $M$, $\chi$ and the quadrupole deviation, which is observed in its absence, is no longer seen, and the posterior distributions are mostly peaked around their injected values. In addition, the quadrupole deviation is also  much better constrained, with the $1\sigma$ limits decreasing by $\sim 90\%$ for $q$ and $\sim 94\%$ for $\epsilon_3$, compared to LIGO GW150914. The two-dimensional contours are also constrained much better, signifying a low posterior volume.

\section{Conclusion}

In this paper, we apply the eikonal approximation to the analysis of real GW data to test the Kerr hypothesis. The ringdown phase of the GW signal contains information about the QNMs emitted by the remnant BH. Measuring these QNMs allows one to gather information about the remnant BH spacetime and the theory of gravity describing BH perturbations. While it has been shown that the final compact object produced by GW150914 is compatible with a Kerr BH, the viability of other alternatives has not been ruled out. We focus here on remnant geometries differing from Kerr as a result of an anomalous quadrupole moment. Its measurement will tell whether the observed data supports the Kerr hypothesis, and whether it is compatible with alternative spacetime geometries as well. 

The main difficulty in performing such ringdown analyses is the calculation of the QNM frequencies in an arbitrary axisymmetric and stationary background spacetime. Although this is a non-trivial task, the eikonal approximation allows one to calculate QNM frequencies without having to explicitly solve the BH perturbation equations, by relating the QNM frequencies to the orbital frequency and Lyapunov exponent of the unstable light ring.

We concentrate our efforts on two spacetime metrics for the remnant - the MN  and the JP metric. We use restricted versions of these metrics, where deviations from Kerr are regulated by one single parameter, the anomalous quadrupole moment. (We stress that our program can also be applied to remnant geometries whose deviations from Kerr are regulated by two or more parameters.)
We use GW150914 data and the eikonal approximation as described in section \ref{sec:methodology} to calculate the posterior distribution for $M, \chi, A_{220}, \phi_{220}$ and the anomalous quadrupole parameter ($q$ or $\epsilon_3$). Our analysis to recover the posterior distributions does not assume deviations from Kerr to be small, and uses full Bayesian methods to calculate the posteriors. The results obtained are shown in Figs. \ref{fig:MNpost} and \ref{fig:JPpost}. While the recovered posteriors are compatible with Kerr, for the JP metric ansatz the remnant mass is  poorly constrained. However, restricting the mass prior by utilizing information from the inspiral-merger analysis, we arrive at bounds on $\epsilon_3$ that are comparable to those from X-ray observations~\cite{Kong:2014wha, Bambi:2015ldr}. The distribution for $\epsilon_3$ is also quite asymmetric and seems to favour positive values. This can be attributed to the dependence of QNM frequencies and damping time  on spin, mass and $\epsilon_3$. We discuss in section \ref{sec:results} that the frequency and damping times calculated from the sampled parameters must simultaneously satisfy the observational bounds from GW150914,  resulting in a skewed distribution.
The recovered posteriors for $q$ and $\epsilon_3$ also differ by $\sim O(100)$. 
As the metrics under consideration are quite involved, we study this feature by assuming a small parameter expansion. We find that the two deviation parameters cannot be mapped to one another, which
explains the different width of their posteriors.

We also consider future (third-generation) detectors capable of detecting GW150914-like signals, such as the Einstein Telescope. Starting from simulated injections we recover posterior distributions using the same approach as for GW150914. The posterior distribution covers a smaller  volume, but the degeneracies between remnant mass, spin and the anomalous quadrupole remain qualitatively unchanged
if one only includes the fundamental mode. However, future detectors will also allow for detecting higher order QNM modes. We find that the inclusion of the (3, 3) mode in the analysis of ET simulated data allows for constraining deviations from the Kerr quadrupole at percent level.

\textit{Acknowledgments:} 
We thank Nicola Franchini, Sebastian H. V\"{o}lkel and Emanuele Berti for providing useful feedback on the manuscript. K.D. acknowledges IISER Thiruvananthapuram for providing high-performance computing resources at HPC Padmanabha. E.B. acknowledges support from the European Union's H2020 ERC Consolidator Grant ``GRavity from Astrophysical to Microscopic Scales'' (Grant No.  GRAMS-815673) and the EU Horizon 2020 Research and Innovation Programme under the Marie Sklodowska-Curie Grant Agreement No. 101007855.


\bibliography{main}

\end{document}